# Optimized Energy Efficient Virtualization and Content Caching in 5G Networks

Ahmed N. Al-Quzweeni[1,2], Ahmed Q. Lawey[1], Taisir E.H. Elgorashi[1], and Jaafar M.H. Elmirghani[1], *Fellow IEEE*

[1] School of Electronic and Electrical Engineering, University of Leeds, LS2 9JT UK.
[2] Computer Technology Department, Faculty of Information Technology, Imam Ja'afar Al-Sadiq University, Baghdad, Iraq

This work was supported by the Engineering and Physical Sciences Research Council (ESPRC), INTERNET (EP/H040536/1), STAR (EP/K016873/1) and TOWS (EP/S016570/1) projects. All data are provided in full in the results section of this paper.

**ABSTRACT** Network function virtualization (NFV) and content caching are two promising technologies that hold great potential for network operators and designers. This paper optimizes the deployment of NFV and content caching in 5G networks and focuses on the associated power consumption savings. In addition, it introduces an approach to combine content caching with NFV in one integrated architecture for energy aware 5G networks. A mixed integer linear programming (MILP) model has been developed to minimize the total power consumption by jointly optimizing the cache size, virtual machine (VM) workload, and the locations of both cache nodes and VMs. The results were investigated under the impact of core network virtual machines (CNVMs) inter-traffic. The result show that the optical line terminal (OLT) access network nodes are the optimum location for content caching and for hosting VMs during busy times of the day whilst IP over WDM core network nodes are the optimum locations for caching and VM placement during off-peak time. Furthermore, the results reveal that a virtualization-only approach is better than a caching-only approach for video streaming services where the virtualization-only approach compared to caching-only approach, achieves a maximum power saving of 7% (average 5%) when no CNVMs inter-traffic is considered and 6% (average 4%) with CNVMs inter-traffic at 10% of the total backhaul traffic. On the other hand, the integrated approach has a maximum power saving of 15% (average 9%) with and without CNVMs inter-traffic compared to the virtualization-only approach, and it achieves a maximum power saving of 21% (average 13%) without CNVMs inter-traffic and 20% (average 12%) when CNVMs inter-traffic is considered compared with the caching-only approach. In order to validate the MILP models and achieve real-time operation in our approaches, a heuristic was developed. The heuristic results are comparable to the MILP model results with and without CNVMs inter-traffic.

**INDEX TERMS** 5G networks, Backhaul, BBU, Energy Efficiency, Fronthaul, IP over WDM, Network Function Virtualization, NFV.

## I. INTRODUCTION

In recent years, the appetite for multimedia services has witnessed a tremendous increase resulting in a Compound Annual Growth Rate (CAGR) in mobile traffic of 46% reaching 48.3 exabytes per month in 2021 [1]. This is further driven by a number of factors such as the emergence of data-hungry applications and the Internet of Things (IoT) [2]. However, the current mobile network may not be able to tackle the growth in traffic volume and the user demands without consuming unnecessarily large energy resources [3]. Compared to the current mobile network, the next generation of mobile networks, namely 5G and beyond are expected to achieve 10 fold growth in energy-efficiency and 1000 times increase in capacity [4, 5]. The 5G networks, which became operational in 2020 [6], are already diverse networks that use a range of multiple access technologies backed by a significant amount of new spectrum to provide different types of users (eg. IoT, personal, industrial) with different services at high data rate and with latencies less than 1 ms [7]. However, the revolutionary requirements of the envisaged 5G / 6G networks have inspired researchers around the world to look for new methodologies, architectures, and technologies to cope with the 5G / 6G requirements [8], [9] and improve the energy-efficiency.
Caching popular content near users and NFV are promising technologies for the design and implementation of 5G / 6G networks [10]. The efficient deployment of content caching in mobile networks can improve the quality of service (QoS) and reduce the backhaul and core network traffic congestion. In a similar vein, NFV enables better network resource utilization, reduces the operational costs and provides seamless development of new services. Although work has been done on NFV and content caching to improve the Information and Communication Technology (ICT) energy-efficiency, these two technologies have been investigated separately in the literature. The authors of [11] studied the impact of optimizing virtual network function (VNF) placement and traffic routing on the energy-efficiency of telecommunication networks. The authors of [12] introduced a framework for designing an energy-efficient architecture for 5G mobile network function virtualization. They investigated the impact of virtualizing the mobile baseband and core network functions on the energy-efficiency of 5G networks. In [13] we investigated an energy aware framework for network function virtualization in 5G networks and studied the impact of caching the contents on energy optimization in 5G networks.

However, it is beneficial to jointly investigate these two technologies to improve the energy efficiency in 5G networks. Therefore, in this paper we extend our previous work in [14] by jointly considering content caching and NFV for reduced energy consumption in 5G networks. This work evaluates the energy consumption of delivering video over optical-based architectures for 5G networks with virtualized mobile functions and resources. A mixed integer linear programming (MILP) model was developed to jointly optimize the cache size and the location of both caches and VMs. To achieve maximum power saving, the MILP model finds the optimum cache size and the optimum VMs workload at each node at different times of the day for different number of users.
Following the introduction, this paper is organized as follows: Section II introduces content caching and NFV in 5G Networks. Section III provides the new MILP model developed, Section IV gives the MILP parameters and describes the new results obtained. Section V discusses the new Energy-Efficient Virtualization and Caching (EEVIRandCa) Heuristic developed and compares its results to those of the MILP thus providing verification and algorithms for real time implementation of the techniques proposed in this paper. The conclusions are finally given in Section VI.

## II. Content caching and NFV in 5G networks

NFV and content caching are promising technologies for 5G networks [12], [15] - [17] where caching the content and user data processing close to the end users result in shorter paths to the content and low traffic induced power consumption. However, this approach requires more equipment for caching the content and hosting VMs which eventually increases the equipment induced power consumption. Therefore, the cache size, VM utilization of nodes, and the location of both cache nodes and VMs are considered in this work. This evaluation minimizes the total power consumption by optimizing the cache size, VM utilization of nodes and the location of caches and VMs.

## III. MILP model

This section introduces the MILP model developed to minimize the total power consumption attributed to VM hosting servers (data processing by VMs), cache nodes, and traffic flow through the network.
**Fig. 1** illustrates the proposed system architecture modelled by the MILP framework developed. In this architecture, a video on demand content caching service is integrated with the virtualization architecture proposed in our previous work in [14]. The architecture in this work consists of three layers: IP over WDM network, wired optical access network represented by a passive optical network (PON), and mobile radio access network (RAN) represented by a set of remote radio head (RRH) nodes. As in the previous work, two types of VMs are proposed. The first type carries out the mobile core network functions (CNVM), whilst the second is in charge of the base band unit (BBU) functions (BBUVM). The video server location is restricted to one of the IP over WDM nodes, whilst the content caches and the two types of VMs (CNVM and BBUVM) can be anywhere (at ONU, OLT, and IP over WDM nodes) in the network. Accordingly, four types of traffic flow in the network: traffic from video servers, traffic from content caches, traffic from CNVMs and traffic from BBUVMs. The BBUVM aggregates the traffic from CNVM, video server, and cache to perform baseband processing and transmits the processed traffic to the RRH node. According to [1] around 80% of the total consumer Internet traffic will be IP video traffic by 2021 and beyond. Therefore, the traffic from video streaming servers together with the traffic from cache nodes are considered as 80% of the total download traffic towards RRH nodes, whilst the traffic from CNVMs makes the remaining 20% of the total traffic.
The developed MILP model aims to investigate the additional power savings that can be achieved through combined caching and NFV in 5G / 6G compared to the virtualization only and caching only approaches. The model also analyses the impact of integration of NFV and content caching on the total power consumption with variation in the total number of network users over the time of the day.
The model declares a number of indices, parameters, and variables. These are listed in Tables I, II, III respectively.

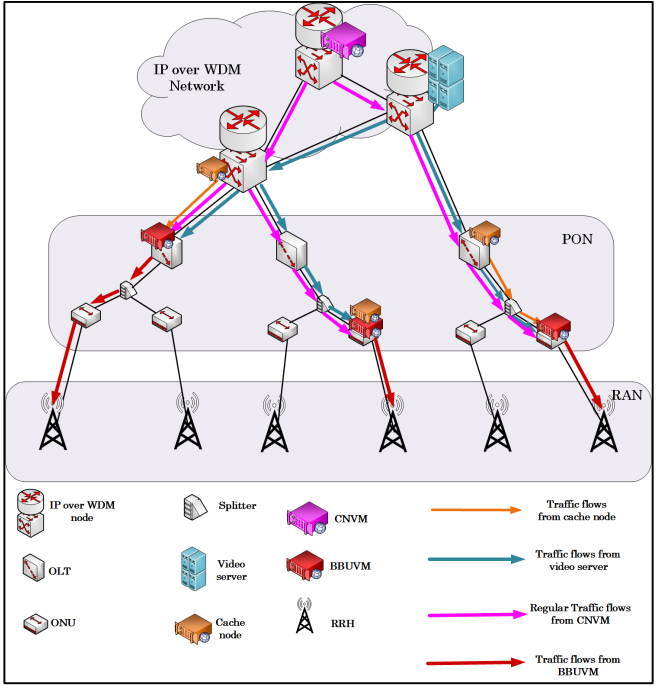

**FIGURE 1** Proposed System Architecture

TABLE I
MILP MODEL INDICES

| Indices | Definition |
|---|---|
| x, y | Indices of any two nodes in the network |
| m, n | Indices of any two nodes in the physical layer of the IP over WDM network |
| i, j | Indices of any two nodes in the IP layer of the IP over WDM network. |
| r | Index of RRH node |
| h, u, p, q | Indices of the nodes where the VM or cache can be hosted |

TABLE II
MILP MODEL PARAMETERS

| Parameters | Definition |
|---|---|
| R | Set of RRH nodes |
| U | Set of ONU nodes |
| L | Set of OLT nodes |
| N | Set of IP over WDM nodes |
| T | Set of all nodes (RRH, ONU, OLT, and IP over WDM nodes) |
| $N_m^N$ | Set of neighbors of node m in the IP over WDM network, $\forall\ m \in N$ |
| $T_x^N$ | Set of neighbors of node x, $\forall\ x \in T$ |
| H | Set of nodes where the VM or cache can be placed (ONU, OLT, and IP over WDM nodes) |
| K | Set of Linearization coefficients |
| l | Line coding rate (bits per sample) |
| y | Number of MIMO layers |
| q | Number of bits used in QAM modulation |
| a | Number of antennas in a cell |
| $c^p$ | CPRI link data rate |
| $\Psi^\chi$ | Maximum BBU workload needed for fully loaded RRH (GOPS); calculated as: $30 \cdot a + 10 \cdot a^2 + 20 \cdot q \cdot l \cdot y$ |
| $\Psi^S$ | Server CPU maximum workload (GOPS) |
| $\Psi^C$ | Workload needed for hosting one CNVM (GOPS) |
| $\rho_r$ | Number of mobile users connected to RRH node r |
| n | Maximum number of resource blocks per cell (per RRH node) |
| $p^b$ | Physical resource block per user |
| $\lambda_r^R$ | RRH node r traffic demand (Gbps); calculated as: $[(p^b/n) \cdot c^p \cdot \rho_r]$, where $r \in R$ |
| $\lambda_r^V$ | Video streaming traffic to RRH node r |
| $\lambda_r^G$ | Regular traffic to the RRH node r |
| μ | Large number |
| $\nabla_{p,q}$ | Intra-traffic between core network VMs (CNVM) at nodes p, and q (Gbps) |
| α | The ratio of the backhaul to the fronthaul traffic (unitless) |
| $\Omega^U$ | ONU maximum power consumption (W) |
| $\Omega^L$ | OLT maximum power consumption (W) |
| $\Omega^{Ld}$ | OLT idle power consumption (W) |
| $C^L$ | OLT maximum capacity (Gbps) |
| $C^U$ | ONU maximum capacity (Gbps) |
| $\Omega_x^R$ | Power consumption of the Remote Radio Head (RRH) connected to ONU node x (W) |
| $\Omega^S$ | Server maximum power consumption (W) |
| $\Omega^{Sd}$ | Server idle power consumption (W) |
| $\Omega_h^H$ | Maximum power consumption of hosting VMs at node h (W) |
| $\Omega^C$ | Cache node maximum power consumption (W) |
| $C^C$ | Cache node maximum storage capacity (GB) |
| $\varepsilon^c$ | Cache node energy per stored gigabyte $\varepsilon c = \Omega^C / C^C$ (W/GB) |
| $\varepsilon^s$ | Video streaming server energy per bit (Joule/Gb) |
| $a_k, b_k$ | Linearization coefficients (unitless) |
| β | Large number |
| η | Very small number (unitless) |
| B | Capacity of the wavelength channel (Gbps) |
| w | Number of wavelengths per fiber |
| $\Omega^T$ | Transponder power consumption (W) |
| $\Omega^{RP}$ | Router power consumption per port (W) |
| $\Omega^G$ | Regenerator power consumption (W) |
| $\Omega^E$ | EDFA power consumption (W) |
| $N_{m,n}^G$ | Number of regenerators in the optical link (m, n) |
| S | Maximum span distance between EDFAs (km) |
| $D_{m,n}$ | Distance between node pair (m, n) in the IP over WDM network (km) |
| $A_{m,n}$ | Number of EDFAs between node pair (m, n) calculated as $A_{m,n} = ((D_{m,n}/S) - 1) + 2$ |

TABLE III
MILP MODEL VARIABLES

| Variables | Definition |
|---|---|
| $\lambda_{p,h}^B$ | Traffic from CNVMs in node p to the BBUVMs in node h (Gbps) |
| $\lambda_{h,r}^R$ | Total download traffic from BBUVMs in node h to the RRH node r (Gbps) |
| $\sigma_{h,r}^B$ | Binary indicator, set to 1 if the node h hosts BBUVMs to serve the RRH node r, 0 otherwise |
| $\sigma_h^B$ | Binary indicator, set to 1 if the node h hosts a BBUVM, 0 otherwise |
| $\sigma_{p,h}^E$ | Binary indicator, set to 1 if the node h hosts CNVMs to serve the BBUVMs at hosting node h, 0 otherwise |
| $\sigma_p^E$ | Binary indicator, set to 1 if the hosting node p hosts CNVMs is, 0 otherwise |
| $\psi_{p,q}$ | Binary indicator, set to 1 if two different hosting nodes p and q host CNVMs, 0 otherwise. It is equivalent to the ANDing of the two binary variables ($\sigma E_p, \sigma E_q$). |
| $\sigma_h^\chi$ | Binary indicator, set to 1 if the hosting node h hosts any virtual machine of any type, 0 otherwise. It is equivalent to the ORing of the two binary variables ($\sigma B_h, \sigma E_h$). |
| $\lambda_{p,q}^E$ | Traffic between hosting nodes due to CNVMs communication (Gbps) |
| $\lambda_{s,r}^{S}$ | Video streaming traffic between BBUVM and RRH node r from a video server at node s |
| $\lambda_{u,r}^{C}$ | Traffic between BBUVM and RRH node r from the cache at node u |
| $\sigma_{u,r}^{C}$ | Binary variable, set to 1 if the cache is located at node u to serve the RRH node r |
| $\lambda_{h,r}^{G}$ | Regular traffic from BBUVM in node h to the RRH node r |
| $\lambda_{u,h,r}^{C}$ | Traffic from the cache at node u to the RRH node r passing through the BBUVM in node h |
| $\lambda_{s,h,r}^{S}$ | Video streaming traffic from video server at node s to the node RRH |
| $\lambda_{s,h}^{S}$ | Video streaming traffic from video server at node s to a BBUVM at node h |
| $\sigma_{s}^{S}$ | Binary variable, set to 1 if a video server is attached to the node s |
| $\lambda_{p,q}^{T}$ | Total download traffic from node p to node q (Gbps) |
| $\lambda_{h,r,x,y}^{R}$ | Traffic from hosting node h to RRH node r that traverses the link between the nodes (x, y) in the network in Gb/s |
| $\lambda_{p,q,x,y}^{T}$ | Total traffic from node p to node q that traverses the link between the nodes (x, y) in the network (Gbps) |
| $\Psi_{h}^{B}$ | Total baseband workload at node h (GOPS) |
| $\Psi_{h}^{i}$ | The integer part of the total normalized workload at node h. |
| $\Psi_{h}^{f}$ | The fractional part of the total normalized workload at node h. |
| $\delta_{u}$ | Hit ratio of the cache at node u |
| $\Theta_{u,r}$ | Floating variable equivalent to the multiplication of the binary variable σCur by the cache hit ratio |
| $Z_{u}^{C}$ | Cache size at node u |
| $Z_{u}^{iC}$ | The integer part of the cache size at node u |
| $Z_{u}^{fC}$ | The fractional part of the cache size at node u |
| $W_{i,j}$ | Number of wavelength channels in the virtual link (i, j) |
| $W_{i,j,m,n}$ | Number of wavelength channels in the virtual link (i, j) that traverse the physical link (m, n) |
| $f_{m,n}$ | Number of fibres in the physical link (m, n) |
| $W_{m,n}$ | Total number of wavelengths in the physical link (m, n) |
| $\Lambda_{m}$ | Number of aggregation ports of the router at node m |

The total power consumption composed of:
1) IP over WDM network power consumption composed of:
a) Router ports power consumption

$$\Omega^{RP} \cdot \left( \sum_{m \in N} \Lambda_m + \sum_{m \in N} \sum_{n \in N_m^N} W_{m,n} \right) \quad (1)$$

b) Transponders power consumption
$$\Omega^{T} \cdot \sum_{m \in N} \sum_{n \in N_m^N} W_{m,n} \quad (2)$$

c) EDFAs power consumption
$$\Omega^{E} \cdot \sum_{m \in N} \sum_{n \in N_m^N} A_{m,n} \cdot f_{m,n} \quad (3)$$

d) Regenerators power consumption
$$\Omega^{G} \cdot \sum_{m \in N} \sum_{n \in N_m^N} N_{m,n}^{G} \cdot W_{m,n} \quad (4)$$

2) RRHs and ONUs power consumption

$$\sum_{x \in U} \left[ \Omega_x^R + \frac{\Omega^U}{C^U} \cdot \left( \sum_{h \in H} \sum_{r \in R} \sum_{y \in T_x^N} \lambda_{h,r,x,y}^R \right. \right.$$
$$\left. \left. + \sum_{p \in H} \sum_{q \in H: p \neq q} \sum_{y \in T_x^N \cap H} \lambda_{p,q,x,y}^T \right) \right] \quad (5)$$

3) OLTs power consumption

$$\sum_{x \in L} \left[ \Omega^{Ld} + \frac{\Omega^L - \Omega^{Ld}}{C^L} \cdot \left( \sum_{h \in H} \sum_{r \in R} \sum_{y \in T_x^N} \lambda_{h,r,x,y}^R \right. \right.$$
$$\left. \left. + \sum_{p \in H} \sum_{q \in H: p \neq q} \sum_{y \in T_x^N \cap H} \lambda_{p,q,x,y}^T \right) \right] \quad (6)$$

4) VM servers power consumption
$$\sum_{h \in H} \left( \Omega^{Sd} \cdot \left( \Psi_h^i + \sigma_h^\chi \right) + \Psi_h^f \cdot \left( \Omega^S - \Omega^{Sd} \right) \right) \quad (7)$$

5) Video server power consumption
$$\varepsilon^S \cdot \sum_{s \in N} \sum_{h \in H} \sum_{r \in R} \left( \alpha \cdot \lambda_{s,h,r}^S \right) \quad (8)$$

6) Caching nodes power consumption
$$\sum_{u \in H} \left( \varepsilon^C \cdot C^C \cdot (Z_u^{iC}/100) \right) \quad (9)$$

The model objective is to minimize the total power consumption:

Minimize:

$$\Omega^{RP} \cdot \left( \sum_{m \in N} \Lambda_m + \sum_{m \in N} \sum_{n \in N_m^N} W_{m,n} \right)$$

$$+ \left( \Omega^T \cdot \sum_{m \in N} \sum_{n \in N_m^N} W_{m,n} \right)$$

$$+ \left( \Omega^E \cdot \sum_{m \in N} \sum_{n \in N_m^N} A_{m,n} \cdot f_{m,n} \right)$$

$$+ \left( \Omega^G \cdot \sum_{m \in N} \sum_{n \in N_m^N} N_{m,n}^G \cdot W_{m,n} \right) +$$

$$\sum_{x \in U} \left[ \Omega_x^R + \frac{\Omega^U}{C^U} \cdot \left( \sum_{h \in H} \sum_{r \in R} \sum_{y \in T_x^N} \lambda R_{x,y}^{h,r} \right. \right.$$

$$\left. \left. + \sum_{p \in H} \sum_{q \in H: p \neq q} \sum_{y \in T_x^N \cap H} \lambda T_{x,y}^{p,q} \right) \right] + \quad (10)$$

$$\sum_{x \in L} \left[ \Omega^{Ld} + \frac{\Omega^L - \Omega^{Ld}}{C^L} \cdot \left( \sum_{h \in H} \sum_{r \in R} \sum_{y \in T_x^N} \lambda_{h,r,x,y}^R \right. \right.$$

$$\left. \left. + \sum_{p \in H} \sum_{q \in H: p \neq q} \sum_{y \in T_x^N \cap H} \lambda_{p,q,x,y}^T \right) \right] +$$

$$\left( \sum_{h \in H} \left( \Omega^{Sd} \cdot \left( \Psi_h^i + \sigma_h^\chi \right) + \Psi_h^f \cdot (\Omega^S - \Omega^{Sd}) \right) \right)$$

$$+ \left( \varepsilon^S \cdot \sum_{s \in N} \sum_{h \in H} \sum_{r \in RRH} \left( \alpha \cdot \lambda_{s,h,r}^S \right) \right)$$

$$+ \left( \sum_{u \in H} \left( \varepsilon^C \cdot C^C \cdot (Z_u^{iC}/100) \right) \right)$$

Subjected to the following constraints:

1. Traffic from CNVMs to BBUVMs

$$\sum_{p \in H} \lambda_{p,h}^B = \alpha \cdot \sum_{r \in R} \lambda_{h,r}^R \quad , \forall h \in H \quad (11)$$

2. Regular traffic to RRH nodes

$$\lambda_r^G = \sum_{h \in H} \lambda_{h,r}^G \quad , \forall r \in RRH \quad (12)$$

Constraint (11) represents the regular traffic from CNVMs to the BBUVMs in node *h* where $\alpha$ is a unitless quantity which represents the ratio of backhaul to fronthaul traffic. Note that this enables BBUVMs to receive traffic from more than a single CNVM, which may occur for example in network slicing. Constraint (12) represents the regular traffic to RRH nodes from all BBUVMs at different nodes. This enables a RRH to receive traffic from more than a single BBUVM.

3. Served RRH nodes and the location of BBUVMs

$$\beta \cdot \lambda_{h,r}^G \geq \sigma_{h,r}^B \quad , \forall r \in R, \forall h \in H \quad (13)$$

$$\lambda_{h,r}^G \leq \beta \cdot \sigma_{h,r}^B \quad , \forall r \in R, \forall h \in H \quad (14)$$

$$\beta \cdot \lambda_{h,r}^G \geq \sigma_{h,r}^B \quad , \forall r \in R, \forall h \in H \quad (15)$$

$$\sum_{h \in H} \lambda_{h,r}^G \leq \beta \cdot \sigma_h^B \quad , \forall r \in R \quad (16)$$

Constraints (13) and (14) ensure that the RRH node *r* is served by the BBUVM located at h. constraints (15) and (16) determine the location of BBUVM; $\beta$ is a large enough number to ensure that $\sigma_{h,r}^B$ and $\sigma_h^B$ are equal to 1 when $\sum_{h \in H} \lambda_{h,r}^G > 0$. In constraint (15) there are two possibilities for the value of ($\sum_{h \in H} \lambda_{h,r}^G$) which are either zero (no traffic from h to r) or greater than zero (there is traffic from h to r). When the value is zero, the left-hand side of the inequality ($\beta \cdot \sum_{h \in H} \lambda_{h,r}^G$) should be zero and this sets the value of $\sigma_h^B$ to zero. In the second case when the value of ($\sum_{h \in H} \lambda_{h,r}^G$) is greater than zero, the left-hand side of the inequality ($\beta \cdot \sum_{h \in H} \lambda_{h,r}^G$) will be much greater than 1 because of the large value of $\beta$. In this case; the value of $\sigma_h^B$ may be set to 1 or zero. In the same way constraint (16) sets the value of $\sigma_h^B$. TABLE IV illustrates the operation of constrains (15) and (16).

TABLE IV
BBUVM CONSTRAINTS OPERATION

| Input | Constraint | Outcome | $\sigma_h^B$ | Value of $\sigma_h^B$ that satisfies both constraints |
|---|---|---|---|---|
| $\sum_{\forall r \in R} \lambda_{h,r}^R > 0$ | $\beta \cdot \sum_{\forall r \in R} \lambda_{h,r}^R \geq \sigma_h^B$ | $\beta \cdot \sum_{\forall r \in R} \lambda_{h,r}^R \gg 1$ | 0 or 1 | 1 |
| | $\sum_{\forall r \in R} \lambda_{h,r}^R \leq \beta \cdot \sigma_h^B$ | $\beta \cdot \sigma_h^B \gg 1$ | 1 | |

| $\sum_{\forall r \in R} \lambda_{h,r}^R = 0$ | $\beta \cdot \sum_{\forall r \in R} \lambda_{h,r}^R \geq \sigma_h^B$ | $\beta \cdot \sum_{\forall r \in R} \lambda_{h,r}^R = 0$ | 0 | 0 |
|---|---|---|---|---|
| | $\sum_{\forall r \in R} \lambda_{h,r}^R \leq \beta \cdot \sigma_h^B$ | $\beta \cdot \sigma_h^B = 0$ | 0 or 1 | |

4. Location of CNVMs

$$\beta \cdot \lambda_{p,h}^B \geq \sigma_{p,h}^E \ , \forall p, q \in H, p \neq q \quad (17)$$

$$\lambda_{p,h}^B \leq \beta \cdot \sigma_{p,h}^E \ , \forall p, q \in H, p \neq q \quad (18)$$

$$\beta \cdot \sum_{h \in H} \lambda_{p,h}^B \geq \sigma_p^E \ , \forall p \in H \quad (19)$$

$$\sum_{h \in H} \lambda_{p,h}^B \leq \beta \cdot \sigma_p^E \ , \forall p \in H \quad (20)$$

$$\psi_{p,q} \leq \sigma_p^E \ , \forall p, q \in H, p \neq q \quad (21)$$

$$\psi_{p,q} \leq \sigma_q^E \ , \forall p, q \in H, p \neq q \quad (22)$$

$$\psi_{p,q} \geq \sigma_p^E + \sigma_q^E - 1 \ , \forall p, q \in H, p \neq q \quad (23)$$

Constraints (17) and (18) ensure that the BBUVMs at node h are served by CNVMs that are located at the node p. Constraints (19) and (20) determine the location of CNVMs by setting the binary variable $\sigma_p^E$ to 1 if there is a CNVM at node p. Constraints (21), (22) and (23) ensure that the CNVMs communicate with each other if they are located at different nodes *p* and *q*. This is equivalent to the logical operation $\psi_{p,q} = \sigma_p^E \ AND \ \sigma_q^E$.

5. CNVMs inter-traffic

$$\lambda E_{p,q} = \nabla_{p,q} \cdot \psi_{p,q} \ , \forall p, q \in H: p \neq q \quad (24)$$

6. Hosting a VM of any type (BBUVM or CNVM)

$$\sigma_h^\chi \leq \sigma_h^B + \sigma_h^E \ , \forall h \in H \quad (25)$$

$$\sigma_h^\chi \geq \sigma_h^B \ , \forall h \in H \quad (26)$$

$$\sigma_h^\chi \geq \sigma_h^E \ , \forall h \in H \quad (27)$$

Constraint (24) represents the traffic between CNVMs at nodes p and q. Constraints (25), (26) and (27) determine if the node h hosts any VM of any type (BBUVM or CNVM).

It is equivalent to the logical operation ($\sigma \chi_h = \sigma E_v \ OR \ \sigma E_a$).

7. Video streaming traffic to RRH nodes

$$\sum_{n \in N} \lambda_{n,r}^S = \lambda_r^V - \sum_{u \in H} \lambda_{u,r}^C \ , \forall r \in R \quad (28)$$

8. Request for video caching

$$\sum_{u \in H} \sigma_{u,r}^C \leq 1 \ , \forall r \in R \quad (29)$$

9. Video (contents) server location

$$\beta \cdot \sum_{r \in R} \lambda_{n,r}^S \geq \sigma_n^S \ , \forall n \in N \quad (30)$$

$$\beta \cdot \sum_{r \in R} \lambda_{n,r}^S \geq \sigma_n^S \ , \forall n \in N \quad (31)$$

$$\sum_{n \in N} \sigma_n^S = 1 \ , \forall n \in N \quad (32)$$

Constraint (28) determines the video streaming traffic sourced by both video (contents) server and cache nodes. Constraint (29) determines whether the RRH node r is served by a cache at node h. Constraints (30) and (31) determine the location of the video (contents) server, whilst constraint (32) ensures it is located at one IP over WDM node.

10. Cache node hit ratio

$$\Theta_{u,r} \leq \sigma_{u,r}^C \ , \forall u \in H, \forall r \in R \quad (33)$$

$$\Theta_{u,r} \leq \delta_u \ , \forall u \in H, \forall r \in R \quad (34)$$

$$\Theta_{u,r} \geq \delta_u - (1 - \sigma_{u,r}^C) \ , \forall u \in H, \forall r \in R \quad (35)$$

$$\Theta_{u,r} \geq 0 \ , \forall u \in H, \forall r \in R \quad (36)$$

$$\delta_u \leq 1 \ , \forall u \in H, \forall r \in R \quad (37)$$

Constraints (33-37) determine the cache hit ratio at any node *h*, where constraints (33-35) are equivalent to the multiplication of the hit ratio by the binary variable $\sigma_{u,r}^C$; whilst constraints (36) and (37) ensure that the hit ratio does not go beyond 1 and is not less than 0. For instance, when the value of $\sigma_{u,r}^C$ is 1, the value of $\Theta_{u,r}$ in constraint (33) takes any value between 0 and 1 including the value of the hit ratio $\delta_u$, whilst in constraint (34) the value of $\Theta_{u,r}$ will be equal to or less than the hit ratio $\delta_u$. In constraint (35) the value of $\Theta_{u,r}$ is equal to or greater than the hit ratio $\delta_u$. In all three cases, the value of the hit ratio $\delta_u$ is the only value for $\Theta_{u,r}$ which satisfies the three constraints. In the same way, when the value of $\sigma_{u,r}^C$ is equal to zero, the value of $\Theta_{u,r}$ is zero. TABLE V illustrates the operation of the constraints (33-37).

TABLE V
CACHE NODE HIT RATIO CONSTRAINTS OPERATION

| Input | Constraint | Outcome | Value of $\Theta_{u,r}$ that satisfies all constraints |
|---|---|---|---|
| $\sigma_{u,r}^C = 1$ | $\Theta_{u,r} \leq \sigma_{u,r}^C$ | any value between 0 and 1 | $\delta_u$ |
| | $\Theta_{u,r} \leq \delta_u$ | $\Theta_{u,r} = \delta_u$ or $\Theta_{u,r} < \delta_u$ | |
| | $\Theta_{u,r} \geq \delta_u \cdot (1 - \sigma_{u,r}^C)$ | $\Theta_{u,r} = \delta_u$ or $\Theta_{u,r} > \delta_u$ | |
| $\sigma_{u,r}^C = 0$ | $\Theta_{u,r} \leq \sigma_{u,r}^C$ | $\Theta_{u,r} = 0$ | 0 |
| | $\Theta_{u,r} \leq \delta_u$ | $\Theta_{u,r} = \delta_u$ or any value $< \delta_u$ | |
| | $\Theta_{u,r} \geq \delta_u \cdot (1 - \sigma_{u,r}^C)$ | $\delta_u = 0$ or any positive number | |

11. Traffic from content cache to the RRH nodes
$$\lambda_{u,r}^C = \Theta_{ur} \cdot \lambda V_r , \forall u \in H, \forall r \in R \quad (38)$$

12. Size of content cache node
$$Z_u^C \geq \delta_u \cdot a_k + b_k , \forall u \in H, \forall k \in K \quad (39)$$

$$Z_u^C = Z_u^{iC} + Z_u^{fC} , \forall u \in H \quad (40)$$

Constraint (38) determines the amount of traffic from the cache node to the RRH based on the cache hit ratio. Constraint (39) calculates the cache size based on the hit ratio using a piecewise linear approximation. Constraint (40) rounds down the cache size (ceiling) to the nearest integer.

13. Cache traffic that passes through BBUVM
$$\lambda_{u,h,r}^C \leq \mu \cdot \sigma_{h,r}^B , \forall u, h \in H, \forall r \in R \quad (41)$$

$$\lambda_{u,h,r}^C \leq \lambda_{u,r}^C , \forall u, h \in H, \forall r \in R \quad (42)$$

$$\lambda_{u,h,r}^C \geq \lambda_{u,r}^C - \mu \cdot (1 - \sigma_{h,r}^B),$$
$$\forall u, h \in H, \forall r \in R \quad (43)$$

$$\lambda_{u,h,r}^C \geq 0 , \forall u, h \in H, \forall r \in R \quad (44)$$

Constraints (41) – (44) ensure that the traffic from the cache node u to the RRH node r passes through the BBUVMs that serve the RRH node r where μ is a large number. The operation of constraints (41) – (43) is illustrated in TABLE VI.

TABLE VI
OPERATION OF CONSTRAINTS (41) – (44)

| Input | Constraint | Outcome | Value of $\lambda_{u,h,r}^C$ that satisfies all constraints |
|---|---|---|---|
| $\sigma_{h,r}^B = 1$ | $\lambda_{u,h,r}^C \leq \mu \cdot \sigma_{h,r}^B$ | any value between 0 and μ | $\lambda_{u,r}^C$ |
| | $\lambda_{u,h,r}^C \leq \lambda_{u,r}^C$ | $\lambda_{u,h,r}^C = \lambda_{u,r}^C$ or $\lambda_{u,h,r}^C < \lambda_{u,r}^C$ | |
| | $\lambda_{u,h,r}^C \geq \lambda_{u,r}^C - \mu \cdot (1 - \sigma_{h,r}^B)$ | $\lambda_{u,h,r}^C = \lambda_{u,r}^C$ or $\lambda_{u,h,r}^C > \lambda_{u,r}^C$ | |
| $\sigma_{h,r}^B = 0$ | $\Theta_{u,r} \leq \sigma_{u,r}^C$ | $\lambda_{u,h,r}^C = 0$ | 0 |
| | $\Theta_{u,r} \leq \delta_u$ | $\lambda_{u,h,r}^C = \lambda_{u,r}^C$ or any value $< \lambda_{u,r}^C$ | |
| | $\Theta_{u,r} \geq \delta_u \cdot (1 - \sigma_{u,r}^C)$ | $\lambda_{u,h,r}^C = 0$ or any positive number | |

14. Video (content) server traffic which passes through BBUVMs

$$\lambda_{n,h,r}^S \leq \mu \cdot \sigma_{h,r}^B,$$
$$\forall s \in N, \forall h \in H, \forall r \in R \quad (45)$$

$$\lambda_{n,h,r}^S \leq \lambda_{n,r}^S, \forall s \in N, \forall h \in H, \forall r \in R \quad (46)$$

$$\lambda_{n,h,r}^S \geq \lambda_{n,r}^S - \mu \cdot (1 - \sigma_{h,r}^B),$$
$$\forall s \in N, \forall h \in H, \forall r \in R \quad (47)$$

$$\lambda_{n,h,r}^S \geq 0, \forall s \in N, \forall h \in H, \forall r \in R \quad (48)$$

Constraints (45) – (48) ensure that the traffic from content (video) server $n$ to the RRH node $r$ passes through the BBUVMs that serve the RRH node $r$ where $\mu$ is a large number. The operation of these constraints is the same as the constraints (41) – (44) explained earlier and summarized in TABLE VI.

15. Total download traffic from BBUVMs to RRH nodes

$$\lambda_{h,r}^R = \lambda_{h,r}^G + \sum_{\forall u \in H} \lambda_{u,h,r}^C + \sum_{\forall n \in N} \lambda_{n,h,r}^S$$
$$\forall h \in H, \forall r \in R \quad (49)$$

16. Total traffic between two hosting nodes

$$\lambda_{p,q}^T = \lambda_{p,q}^E + \lambda_{p,q}^B + \alpha \cdot \sum_{r \in R} \lambda_{p,q,r}^C$$
$$\forall p \in U \cup L, \forall q \in H, p \neq q \quad (50)$$

$$\lambda_{p,q}^T = \lambda_{p,q}^E + \lambda_{p,q}^B + \alpha \cdot \left( \sum_{r \in RRH} \lambda_{p,q,r}^C + \sum_{r \in RRH} \lambda_{p,q,r}^S \right)$$
$$\forall p \in N, \forall q \in H, p \neq q \quad (51)$$

Constraint (49) calculates the total download traffic from BBUVMs at node h to the RRH node r which is sourced by three nodes. These are: video server ($\lambda_{n,h,r}^S$), cache node ($\lambda_{u,h,r}^C$), and the regular traffic from CNVMs ($\lambda_{h,r}^G$). Constraints (50) and (51) calculate the total download traffic between any two hosting nodes.

17. Total BBUVM workload at hosting node $h$

$$\Psi_h^B = \left( \left( \sum_{\forall r \in R} \lambda_{h,r}^R \right) / c^p \right) \cdot \Psi^\chi, \forall h \in H \quad (52)$$

18. Total normalized workload at hosting node $h$

$$\Psi_h^i + \Psi_h^f = (\Psi_h^B + \Psi_h^C)/\Psi^S, \forall h \in H \quad (53)$$

19. Hosting node capacity

$$\left( \Omega^{Sd} \cdot (\Psi_h^i + \sigma_h^\chi) + \Psi_h^f \cdot (\Omega^S - \Omega^{Sd}) \right) \leq \Omega_h^H$$
$$\forall h \in H \quad (54)$$

Constraint (52) calculates the total BBU workload needed for the total download traffic towards RRH nodes, whilst constraint (53) determines the fractional and integer parts of the normalized workload. Constraint (54) ensures that the total workload of a VM hosting server does not exceed the total capacity of the hosting node.

20. Flow conservation of total downlink traffic from BBUVMs to RRH nodes

$$\sum_{y \in T_x^N} \lambda_{h,r,x,y}^R - \sum_{y \in T_x^N} \lambda_{h,r,y,x}^R = \begin{cases} \lambda_{h,r}^R & if\ x \\ -\lambda_{h,r}^R & if\ x \\ 0 & other \end{cases} \quad (55)$$

$$\forall h \in H, \forall r \in RRH, \forall x \in TN_x$$

21. Flow conservation of total downlink traffic between two hosting nodes

$$\sum_{y \in T_x^N \cap H} \lambda_{p,q,x,y}^T - \sum_{y \in T_x^N \cap H} \lambda_{p,q,y,x}^T = \begin{cases} \lambda_{p,q}^T \\ -\lambda_{p,q}^T \\ 0 \end{cases} \quad o \quad (56)$$

$$\forall p, q, x \in H : p \neq q$$

22. GPON link constraints

$$\sum_{h \in H} \sum_{r \in R} \sum_{j \in T_i^N \cap L} \lambda_{h,r,i,j}^R + \sum_{p \in H} \sum_{q \in H, q \neq p} \sum_{j \in T_i^N \cap L} \lambda_{p,q,}^T \quad (57)$$

$$\forall i \in U$$

$$\sum_{h \in H} \sum_{r \in R} \sum_{j \in T_i^N \cap N} \lambda R_{i,j}^{h,r} + \sum_{p \in H} \sum_{q \in H, q \neq p} \sum_{j \in T_i^N \cap N} \lambda T_{i,j}^{p,} \quad (58)$$

$$\forall i \in L$$

Constraint (55) represents the flow conservation of total download traffic towards RRH nodes, whilst constraint (56) represents the flow conservation of the total download traffic between two hosting nodes. Constraints (57) and (58) ensure that the download traffic of GPONs does not flow in the opposite direction.

23. Virtual link capacity of IP over WDM network

$$\sum_{p \in H} \sum_{q \in H, q \neq p} \lambda^T_{p,q,i,j} + \sum_{h \in H} \sum_{r \in R} \lambda^R_{h,r,i,j} \leq W_{i,j} \cdot B \quad (59)$$

$\forall i, j \in N, i \neq j.$

24. Flow conservation in the optical layer of IP over WDM network

$$\sum_{n \in N^N_m} W_{i,j,m,n} - \sum_{n \in N^N_m} W_{i,j,n,m} = \begin{cases} W_{i,j} & if\ n = \\ -W_{i,j} & if\ n = \\ 0 & otherw \end{cases} \quad (60)$$

$\forall i, j, m \in N, i \neq j$

Constraint (59) ensures that the total traffic that travers the virtual link (i, j) does not exceed its capacity, in addition it determines the number of wavelength channels that carry the traffic burden of that link. Constraint (60) represents the flow conservation in the optical layer of the IP over WDM network. It ensures that the total expected number of incoming wavelengths for the IP over WDM nodes of the virtual link (i, j) is equal to the total number of outgoing wavelengths of that link.

25. Number of wavelength channels

$$\sum_{i \in N} \sum_{j \in N: i \neq j} W_{i,j,m,n} \leq w \cdot f_{m,n} \quad (61)$$

$\forall m \in N, \forall n \in N^N_m$

$$W_{m,n} = \sum_{i \in N} \sum_{j \in N: i \neq j} W_{i,j,m,n}$$
$\forall m \in N, \forall n \in N^N_m \quad (62)$

26. Number of aggregation ports

$$\Lambda_i = \left( \sum_{j \in L \cap T^N_i} \left( \sum_{p \in H} \sum_{q \in H, q \neq p} \lambda^T_{p,q,i,j} + \sum_{h \in H} \sum_{r \in R} \lambda^R_{h,r,i,j} \right) \right) / B \quad (63)$$

$\forall i \in N$

Constraints (61) and (62) are the constraints of the physical link (m, n). Constraint (61) ensures that the total number of wavelength channels in the logical link (*i*, *j*) that traverses the physical link (m, n) do not exceed the fiber capacity. Constraint (63) determines the number of wavelength channels in the physical link and ensures that it is equal to the total number of wavelength channels in the virtual link traversing that physical link. Constraint (63) determines the required number of aggregation ports in each IP over WDM router.

## IV. MILP model parameters and results

The developed MILP model was run on high performance computing nodes (HPC) provided through a partnership of the most research-intensive universities in the North of England. The standard computation mode that was used to run the MILP model, is one of four computational modes that exist in the used HPC. It provides a cluster of 252 nodes with up to 6048 cores in total and supports 65 TFLOPS peak using IBM's iDataPlex hardware; and includes a high throughput cluster with 1GB RAM per nodes, 2900 cores based on twin nodes with 6-core Westmere processors supported by 1GE connectivity between nodes. IBM ILOG CPLEX (12.7) optimization studio was used as the optimization software package where it uses the simplex algorithm [18] to solve the MILP models.

The developed MILP model considers the network topology shown in Fig. 2. The considered network topology consists of 3 IP over WDM nodes and 6 GPON networks connected in pairs to the IP over WDM network nodes; two GPON networks for each IP over WDM node. Each GPON network consists of one OLT and three ONUs and each ONU is connected to one RRH node. The topology has one video server whose location in the IP over WDM network is optimized by the developed MILP model to minimize the total power consumption. Each RRH node in the network supports a small cell with maximum number of users equal to 10. Each user in the small cell is allocated 5 physical resource blocks (PRB) as the users are assumed to request the same task from the network. The average number of users in the network varies over the time of day according to the network user profile shown in [14]. Therefore, the amount of downlink traffic to each RRH node is influenced by the total number of active users in the small cell where its maximum value is considered less than 10 Gbps. The total download traffic from both video server and cache nodes to the RRH nodes is considered 80% of the total download traffic as explained earlier.

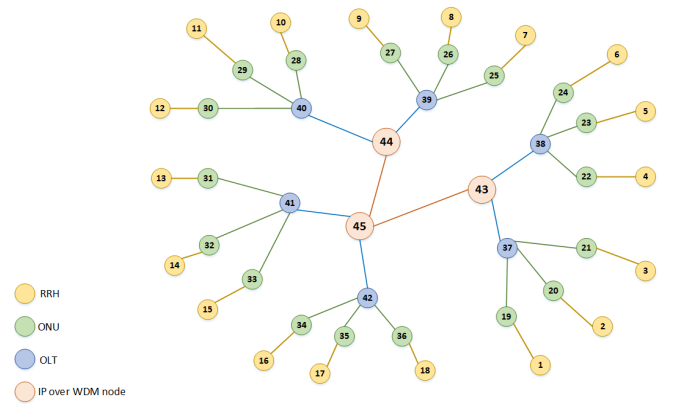

**Figure 2 Network topology used in the evaluation**

Two virtualization approaches were considered; with and without CNVMs inter-traffic. In addition, the content caching approach was considered with variable cache size where the developed MILP model optimized the size of the cache at each node. The input parameters to the developed MILP model are listed in TABLE VII

TABLE VII
MILP MODEL INPUT PARAMETERS

| Parameters | Comments |
|---|---|
| Line coding rate for 8B/10B line coding ($l$) | 10/8 (bit / sample) |
| Number of MIMO layers ($y$) | 2 |
| Number of bits used in QAM modulation for 64 QAM modulation ($q$) | 6 (bits) |
| Number of antennas in a cell ($a$) | 2 |
| Maximum fronthaul (CPRI) data rate for CPRI line rate option 7 ($cp$) | 9.8304 (Gbps) [19] |
| Maximum baseband processing workload needed for fully loaded RRH ($\Psi^x$) given by: $30 \cdot a + 10 \cdot a^2 + 20 \cdot q \cdot l \cdot y$ | 400 (GOPS) |
| Server CPU maximum workload ($\Psi^S$) | 368 (GOPS) [20] |
| Workload needed for hosting one CNVM ($\Psi_h^C$) | 26.17 (GOPS) |
| Number of active users in a small cell ($\rho_r$) | Uniformly distributed (1-10 users) |
| Maximum number of users per cell ($n$) | 10 (users) |
| Number of physical resource blocks per user ($pb$) | 5 (PRB) |
| The ratio of the backhaul to the front haul traffic ($\alpha$) | 0.1344 (unitless) |
| ONU maximum power consumption ($\Omega^U$) | 15 (W) [21] |
| OLT maximum power consumption ($\Omega^L$) | 1940 (W) [22] |
| OLT idle power consumption ($\Omega^{Ld}$) | 60 (W) [22] |
| OLT maximum capacity ($C^L$) | 8600 (Gbps) [22] |
| ONU maximum capacity ($C^U$) | 10 (Gbps) [21] |
| RRH node power consumption ($\Omega_x^R$) | 1140 (W) [23] |
| Hosting server maximum power consumption ($\Omega^S$) | 365 (W) [24] |
| Hosting server idle power consumption ($\Omega^{Sd}$) | 112 (W) [24] |
| Cache node maximum power consumption ($\Omega^C$) | 550 (W) [25] |
| Cache node maximum storage capacity ($C^C$) | 14.4 (TB) [25] |
| Video streaming server energy per bit ($\varepsilon^s$) | 211.1 (Joul/Gb) [26] |
| Capacity of IP over WDM wavelength channel ($B$) | 40 (Gbps) [27-29] |
| Number of wavelengths per fiber in IP over WDM network ($w$) | 32 [27] |
| Transponder power consumption ($\Omega^T$) | 167 (W) [30] |
| Router port power consumption ($\Omega^{RP}$) | 825 (W) [31] |
| Regenerator power consumption ($\Omega^G$) | 334 (W) [31] |
| EDFA power consumption ($\Omega^E$) | 55 (W) [31] |
| Maximum span distance between EDFAs ($S$) | 80 (km) [27, 28] |

As discussed, the MILP model was developed to minimize the total power consumption by optimizing the caches sizes, VM servers' utilization, and the location of both VMs and caches at different node. The results compare the utilization of virtualization and content caching in 5G / 6G individually and investigate the total power consumption of each approach. In addition, the impact of integrating content caching and virtualization is compared with the impact of the deployment of each technology individually. Fig. 3 illustrates the total power consumption of caching-only, virtualization-only (with and without CNVMs inter-traffic), and the integrated (integrated caching and virtualization) approaches for different times of the day, whilst Fig. 4 illustrates the total power consumption for the same approaches for different number of users.

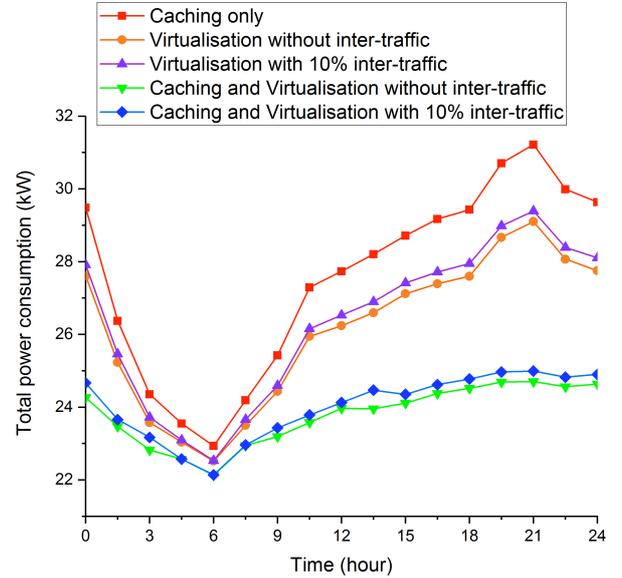

**Figure 3. Total power consumption of different approaches at different times of the day**

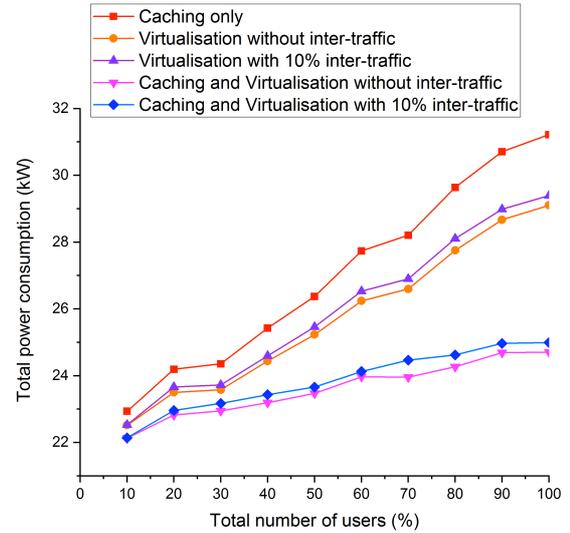

**Figure 4. Total power consumption of different approaches for different number of users**

The caching-only approach has higher power consumption compared to the other approaches whilst the integrated caching and virtualization approaches without CNVMs inter-traffic have the lowest power consumption among all the approaches.

Although the virtualization only approaches (with and without CNVMs inter-traffic) have high power consumption compared to the approach where the caching and virtualization are integrated (integrated approach), they have less power consumption compared to the caching-only approach. This is attributed to the fact that the virtualization-only approach achieves much lower mobile functions power consumption compared to the caching-only approach as shown in Fig. 5. Therefore, the total power consumption of caching-only approach is higher than the virtualization-only approach in spite of its lower video streaming service power consumption compared to the virtualization-only approach as shown in Fig. 6.

power saving of 7% (average 5%) when no CNVMs inter-traffic is considered and 6% (average 4%) with CNVMs inter-traffic of 10% of the total backhaul traffic.

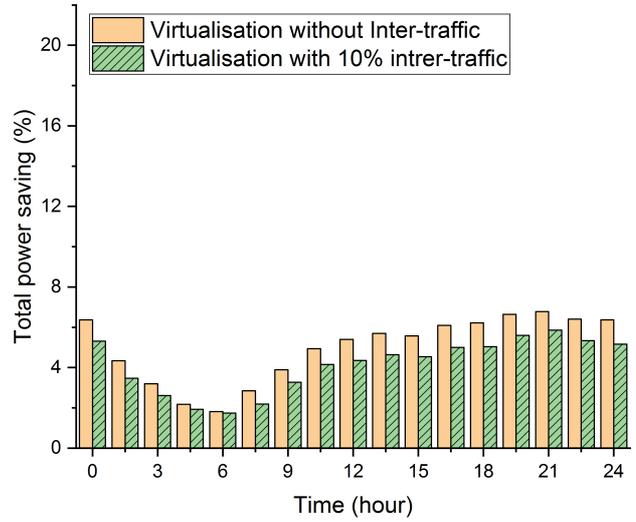

FIGURE 7. A) Power saving of virtualization-only compared with caching-only approach at different times of the day

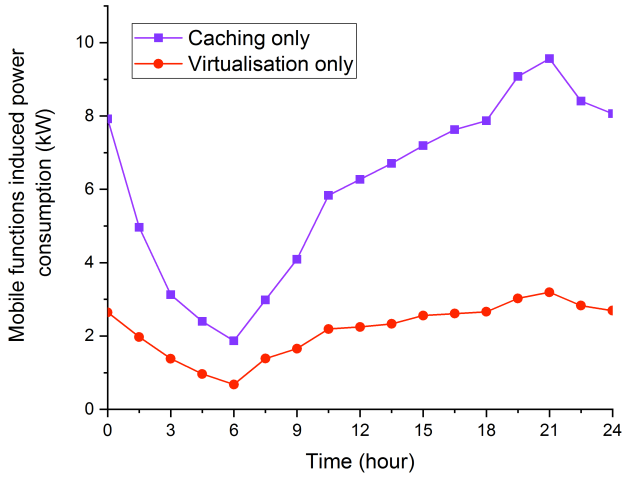

FIGURE 5. Mobile function induced power consumption, of caching and virtualization only approaches

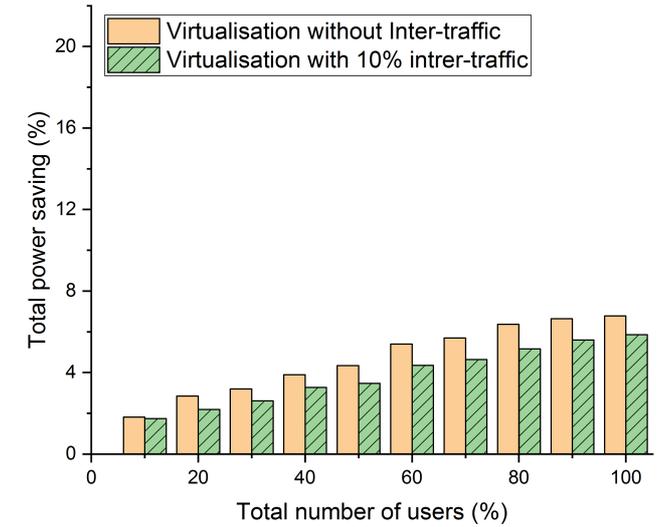

FIGURE 7. B) Power saving of virtualization-only compared with caching-only approach at different number of users

The deployment of virtualization and content caching in one integrated architecture has a great impact on the total power consumption. Fig. 8. illustrates the total power saving of the integrated approach compared with the virtualization-only and caching-only approaches. Compared with the virtualization-only approach, the integrated approach has a maximum total power saving of 15% (average 9%) with and without CNVMs inter-traffic. This is due to the lower video streaming service power consumption of the integrated approach compared to the virtualization only approach (as shown in Fig. 9, where a maximum video streaming service power saving of 95% (average 90%) is achieved by the

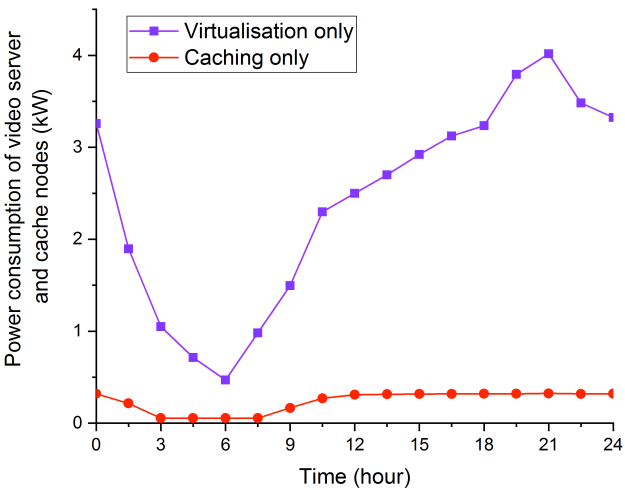

FIGURE 6. Video streaming service power consumption in caching-only and virtualization-only approaches

Fig. 7. illustrates the saving in total power consumption of virtualization-only approach compared to the caching-only approach for different times of the day and different number of users. Compared to the caching-only approach, the virtualization-only approach achieves a maximum total

integrated approach compared with virtualization-only approach.

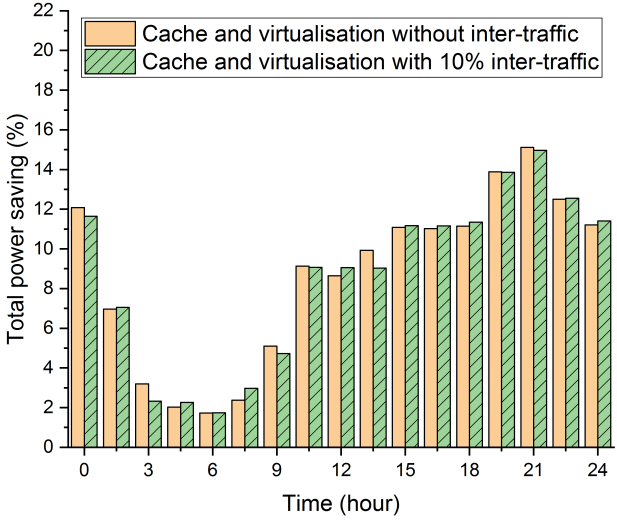

FIGURE 8. A) Power saving of integrated virtualization and caching approach compared to virtualization-only approach at different times of the day

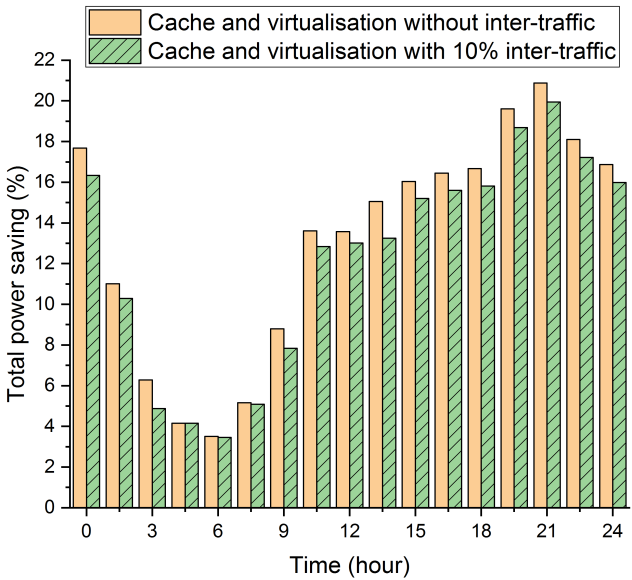

FIGURE 8. B) Power saving of integrated virtualization and caching approach compared to caching-only approach at different times of the day

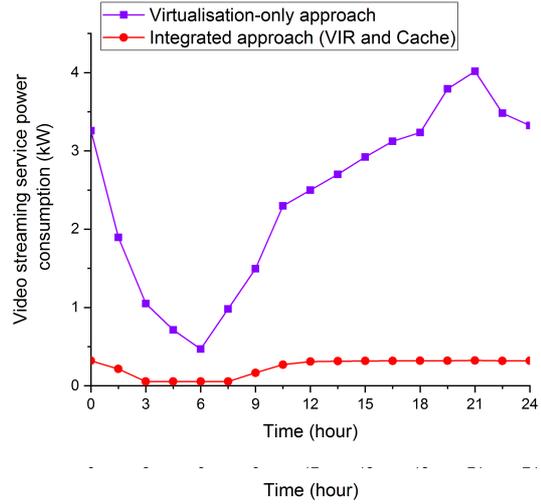

FIGURE 9. Video streaming service power consumption for virtualization-only and integrated approaches at different times of the day

Fig.10 depicts the integrated approach optimum cache utilization of each node at different times of the day when there is no CNVMs inter-traffic and also for CNVMs inter-traffic of 10% of the total backhaul traffic. The optimum cache size at each node varies with the delivered traffic from the node over time. The optimum cache size is relatively high when the total number of users is high during the busy hours of the day and the caches are distributed close to the users. The OLT nodes are highly utilized by caches during the busy times of the day whilst the IP over WDM nodes are utilized when few users are active. During the busy time of the day, there is a large number of active users and therefore the demand for video streaming is high. In this case the integrated approach accommodates a large amount of the video files (objects) at OLT nodes to serve as much as possible users and offload the traffic away from the core network whilst in the virtualization-only approach, the users are served by the video server in the core network.

The integrated approach achieves a maximum total power saving of 21% (average 13%) without CNVMs inter-traffic and 20% (average 12%) when CNVMs inter-traffic is considered compared with the caching-only approach. Additionally, the integrated approach has a maximum mobile function power saving of 65% (average 57%) without CNVMs inter-traffic and 70% (average 58%) when CNVMs inter-traffic is considered at a level of 10% of the total backhaul traffic. The results are in comparison with the caching-only approach. These power savings are driven by the higher mobile functions power consumption in the caching-only approach which is much higher than the integrated approach as shown in Fig. 11.

amount of offloaded backhaul traffic, and the inter-traffic between CNVMs. It is clearly seen in Fig. 12 that in the integrated approach, the OLT nodes are highly utilized by VMs during the busy time of the day to maximize the number of users severed by each VM and minimize the backhaul traffic carried over the core network. When the total number of users is low during off peak times of the day, the IP over WDM nodes are utilized by VMs. In this case the traffic induced power consumption is low compared to the VM power consumption, therefore, the total number of active users is served by a small number of VMs located far from the cells specifically in the IP over WDM network nodes.

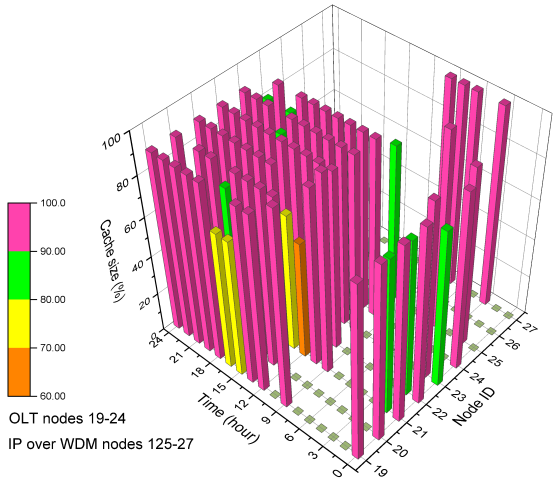

**FIGURE 10 A) Cache utilization of each node at different times of the day without CNVMs inter-traffic**

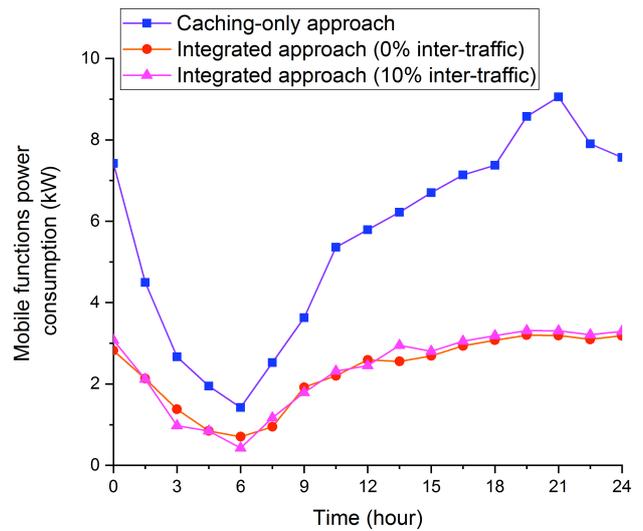

**FIGURE 11 Mobile functions power consumption of caching-only and integrated approaches at different times of the day**

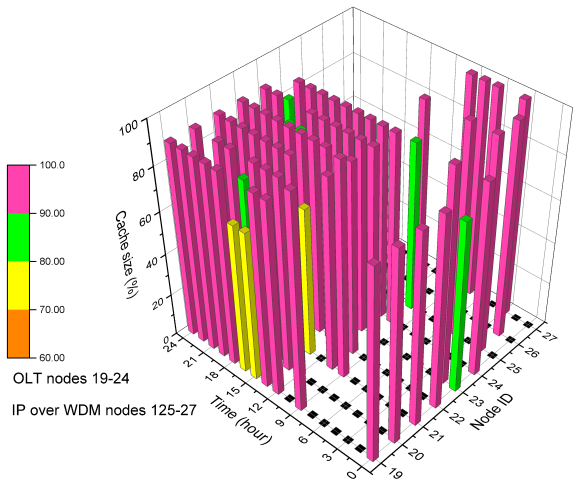

**FIGURE 10 B) Cache utilization of each node at different times of the day with 'CNVMs inter-traffic of 10% of the total backhaul traffic'**

It is worth mentioning that compared to Fig. 5., the integrated approach (Caching and virtualization) in Fig. 11 does not provide improvement over the results of virtualization only (Fig. 5) as the traffic in Fig. 11 does not include video traffic.

Fig. 12 depicts the integrated approach optimum VM (BBUVM and CNVM) utilization of each node at different times of the day when the CNVMs inter-traffic is 0% and 10% of the total backhaul traffic. The optimum utilization of each node varies with the traffic delivered from the node at a certain time. The optimum VM (BBUVM and CNVM) workload at each node is relatively high when the total number of users served by the VM is high during the busy hours of the day where the VMs are distributed close to the small cells. The optimum VM workload and location are mainly driven by: the total number of active users, the

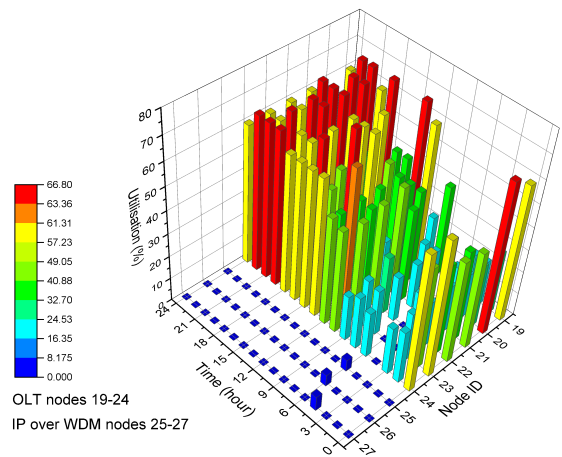

**FIGURE 12 A) VM utilization of each node at different times of the day without CNVMs inter-traffic**

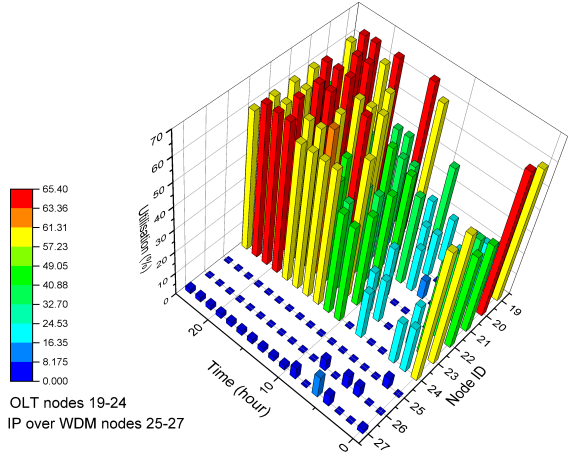

**FIGURE 12 B) VM utilization of each node at different times of the day 'with CNVMs inter-traffic of 10% of backhaul traffic'**

## V. Energy-Efficient Virtualization and Caching (EEVIRandCa) Heuristic Implementation and Results

This section introduces an Energy-Efficient Virtualization and content caching (EEVIRandCa) heuristic approach for real-time implementation of the developed MILP model. The heuristic is also independent of the MILP and therefore provides verification of the MILP and its results. The pseudocode of the EEVIRandCa heuristic is shown in Algorithm 1. Two topologies are considered by the heuristic model; the network to topology $G = (NE, L)$ and the physical topology of the IP over WDM network $G_p = (N, L_p)$, where $NE, L, N, L_p$ are the sets of network elements, link, IP over WDM nodes and the physical links respectively. The network elements $NE$ which are $RRH, ONU, OLT$, and IP over WDM nodes $N$ are captured in a set. Besides the set of nodes, three network elements contribute to the power consumption which are; video streaming servers, cache nodes, and servers that host virtual machines. The total number of active users in each cell ($RRH$ node) determines the total requested download traffic (fronthaul traffic) for each cell ($RRH$), whilst the video streaming traffic and regular traffic account for the total traffic and are determined using Cisco VNI data [1], whilst the cache size of each node is determined based on the optimized hit ratio of each cache node. The EEVIRandCa heuristic starts accommodating BBUVMs in such a way that BBUVMs serve as many RRH requests as possible. The heuristic examines the OLT nodes to determine the closest OLT for each RRH node to accommodate a BBUVM.

To accommodate CNVMs, the EEVIRandCa heuristic sets an initial value for the number of CNVMs in the IP over WDM network. The initial number of CNVMs in the IP over WDM network is increased by one every time the heuristic assesses the IP over WDM network power consumption to determine the optimum number of CNVMs. To determine the highly recommended nodes for accommodating CNVMs, the heuristic builds a sorted list of IP over WDM nodes based on the total number of users connected directly to each node. The first node at the top of the sorted list of IP over WDM nodes is assigned by the heuristic to accommodate the video streaming servers.

Using the same methodology, the EEVIRandCa heuristic determines the location of the caches. It sets an initial value of one cache and increases this value every time it evaluates the IP over WDM network power consumption to obtain the best value for the number of caches in the IP over WDM network. The heuristic calculates the total number of active users and compares it to the maximum network capacity. If the total number of active users is less than 50% of the total network capacity, the EEVIRandCa heuristic examines the IP over WDM network otherwise it examines the OLT nodes to cache the content.

Once the CNVMs, BBUVM, video servers, and cache nodes location are determined, the traffic from these entities is determined and routed towards the RRH nodes. The traffic from CNVMs, video servers and the cache nodes should pass through BBUVMs for BBU processing. In addition, the inter-traffic between CNVMs and total traffic flows in the IP over WDM network are determined.

The EEVIRandCa heuristic obtains the physical graph $G_p = (N, L_p)$ and determines the traffic in each network segment. The IP over WDM network configuration such as the number of fibres, router ports, and the number of EDFAs is then determined and the total power consumption is finally calculated.

---

**Algorithm 1** Energy Efficient Virtualization and Caching (EEVIRandCa) Heuristic

**INPUT:** $G = (NE, L), G_p = (N, L_p)$

**OUTPUT:** VMs location, VMs workloads, VM and Cache distribution, and optimum total power consumption (OTPC)

1: $\forall r \in RRH$ determine number of users and calculate node demand $(r, D_r)$; where $D_r \in D$ /*$D$ is the total demands*/
2: $\forall r \in RRH$ calculate the streaming traffic $S_r$ and regular traffic $L_r$ /*$D_r = S_r + L_r$*/
3: $\forall D_r \in D$ determine BBUVM workload $\Psi_r$
4: $\forall r \in RRH$ find
   $(r, h) = min\,(shortestPath(r, \{h \in NE \cap OLT\})$
5: **if** total workload of $h >> \Psi_r$
6:     host BBUVM in $h$
7:     update workload of $h$
8:     $D_r \in D_{served}$
9:  **end if**
10: x$\forall D_r \notin D_{served}$ find $(r, h) = min(shortestPath(r, \{h \in$

---

```
            N}))) /*where N is the IP over WDM nodes */
11:     host BBUVM in h
12:     update workload of h
13:     D ∈ D_served
14:   Route the fronthaul traffic from BBUVMs to RRH
        nodes
15:   N' ← DESCEND_SORT(N) and set i = 1
16:   Host CNVM in N'(i), N'(i − 1),… N'(1)
17:   ∀ CNVM in n ∈ N' and ∀BBUVM in h ∈ NE
        find (n, h) = min(shortestPath(n, h))
18:   Route the traffic from CNVMs to BBUVMs
19:   ∀ CNVM in n_x, n_y ∈ N'; x ≠ y
        find (n_x, n_y) = min(shortestPath(n_x, n_y))
20:   Route the traffic from CNVMs to BBUVMs
21:   Determine the IP over WDM network configuration
22:  if i = 1
23:    Determine vTPC /* where vTPC is the total power
        consumption due to virtualization */
24:  end if
25:    Determine total power consumption as TPC(i)
26:  if vTPC ≥ TPC(i)
27:     vTPC = TPC(i)
28:     i = i + 1
29:     goto 16
30:  else
31:    virtualization minimum power consumption is vTPC
32:    EXIT
33:  end ifx
34:   N'' ← DESCEND_SORT(N) and set j = 1
35:   host the video server at N(1)
36:   Host cache in N''(i), N''(i − 1),… N''(1)
37:   Route the traffic to RRH through BBUVMs
38:  if j = 1
39:    Determine caTPC /* where caTPC is the total power
        consumption due to virtualization */
40:    EXIT
41:  end if
42:    Determine total power consumption as TPC(j)
43:  if caTPC ≥ TPC(j)
44:     caTPC = TPC(j)
45:     j = j + 1
46:     goto 36
47:  else
48:    caching minimum power consumption is caTPC
49:  end if
50:    minTPC = vTPC + caTPC /*minTPC is the minimum
        power consumption */
```

For fair validation of the MILP model, the network topology in Fig. 2. is used by the EEVIRandCa heuristic. In addition, the parameters considered in the MILP models such as the wireless bandwidth, number of resources blocks per user, and the parameters listed in TABLE VII are considered in the heuristic. The number of users allocated to each cell in the heuristic is the same as in the MILP model to ensure the requested traffic by each RRH node is the same in both approaches.

Fig. 13. compares the total power consumption of the EEVIRandCa heuristic with the power consumption following the optimum solution obtained by the MILP model at different times of the day when there is no CNVMs inter-traffic flows between the VMs. It is clearly seen that the EEVIRandCa heuristic has a higher power consumption compared to the integrated approach MILP model. The difference in power consumption between the two models varies according to the total number of users during the day and is 3.3% max (2% average).

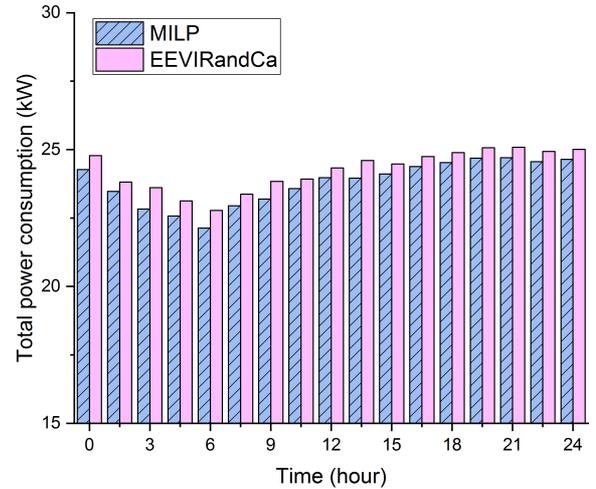

**FIGURE 13. Total power consumption of EEVIRandCa heuristic model compared with the integrated approach MILP model when no CNVMs are considered**

Fig.14. compares the total power consumption of the EEVIRandCa heuristic with the corresponding power consumption when the MILP model is used, when the CNVMs inter-traffic is 10% of the total backhaul traffic recorded at different times of the day. The EEVIRandCa heuristic results are comparable to those of the MILP model.

The difference in power consumption between the two varies during the day and is 2% max (average 1%).

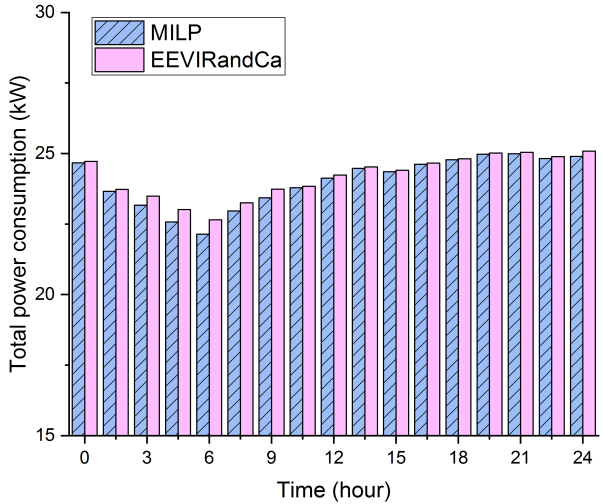

**FIGURE 14. Total power consumption of EEVIRandCa compared to the integrated approach MILP model when CNVMs inter-traffic is 10% of the total backhaul traffic**

## VI. Conclusions

This paper examined NFV and content caching in 5G networks and optimized their use to minimize the associated power consumption. In addition, it introduced an approach that combines content caching with NFV in one integrated architecture for 5G (and beyond) networks. A MILP model was developed to minimize the total power consumption by jointly optimizing the cache size, VM workload, and the locations of both cache nodes and VMs. The results of the developed model were investigated under the impact of CNVMs inter-traffic. The result show that the OLT nodes are the optimum location for content caching and hosting VMs during busy times of the day whilst IP over WDM nodes are the optimum locations for caching and virtualization during off-peak time. Therefore, the appropriate nodes can be activated at different times of the day for content caching and to host VMs.

The results show that the virtualization-only approach is better than caching-only approach for video streaming services where the virtualization-only approach compared to caching-only approach, achieves a maximum power saving of 7% (average 5%) when no CNVMs inter-traffic is considered and 6% (average 4%) with CNVMs inter-traffic at 10% of the total backhaul traffic. On the other hand, the integrated approach has a maximum power saving of 15% (average 9%) with and without CNVMs inter-traffic compared to the virtualization-only approach, and it achieves a maximum power saving of 21% (average 13%) without CNVMs inter-traffic and 20% (average 12%) when CNVMs inter-traffic is considered compared with the caching-only approach.

In order to validate the MILP model and achieve real-time operation of our approaches, a heuristic (EEVIRandCa) was developed. The heuristic results are comparable to the results of the MILP model with and without CNVMs inter-traffic; with a maximum difference of 3.3% over the range of scenarios considered.

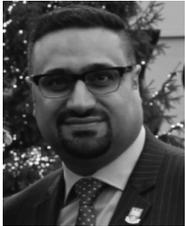

**Ahmed N. Al-Quzweeni** received the B.Sc. and M.Sc. degrees in computer engineering from Nahrain University, Baghdad, Iraq, in 2001 and 2004, respectively, and the Ph.D. degree in communication networks from the University of Leeds, U.K., in 2019. From 2005 to 2009, he was a Mobile Core Network Senior Engineer, where he was involved in short message system, intelligent network, PSTN, and billing systems. From 2009 to 2014, he was an Assistant Lecturer with the Department of Computer Communication, Al-Mansour University College, Baghdad. He was a Team Leader with ZTE Corporation for Telecommunication, Iraq. He is currently with Imam Ja'afar Al-Sadiq University, Baghdad, Iraq. His current research interests include energy efficiency in optical and wireless networks, NFV, mobile networks, 5G networks, content caching, cloud computing, and the Internet of Things.

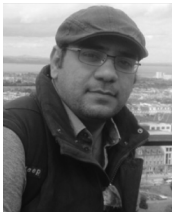

**Ahmed Q. Lawey** (Associate Member, IEEE) received the B.S. (Hons.) and M.Sc. (Hons.) degrees in computer engineering from the University of Al-Nahrain, Iraq, in 2002 and 2005, respectively, and the Ph.D. degree in communication networks from the University of Leeds, U.K., in 2015. From 2005 to 2010, he was a Core Network Engineer with ZTE Corporation for Telecommunication, Iraq. He is currently a Lecturer in communication networks with the School of Electronic and Electrical Engineer, University of Leeds. His current research interests include energy efficiency in optical and wireless networks, big data, cloud computing, and the Internet of Things.

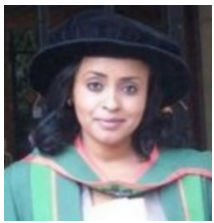

**Taisir E. H. EL-Gorashi** received the B.S. degree (first-class Hons.) in Electrical and Electronic Engineering from the University of Khartoum, Khartoum, Sudan, in 2004, the M.Sc. degree (with distinction) in Photonic and Communication Systems from the University of Wales, Swansea, UK, in 2005, and the PhD degree in Optical Networking from the University of Leeds, Leeds, UK, in 2010. She is currently a Lecturer in optical networks in the School of Electronic and Electrical Engineering, University of Leeds. Previously, she held a Postdoctoral Research post at the University of Leeds (2010– 2014), where she focused on the energy efficiency of optical networks investigating the use of renewable energy in core networks, green IP over WDM networks with datacenters, energy efficient physical topology design, energy efficiency of content distribution networks, distributed cloud computing, network virtualization and big data. In 2012, she was a BT Research Fellow, where she developed energy efficient hybrid wireless-optical broadband access networks and explored the dynamics of TV viewing behavior and program popularity. The energy efficiency techniques developed during her postdoctoral research contributed 3 out of the 8 carefully chosen core network energy efficiency improvement measures recommended by the GreenTouch consortium for every operator network worldwide. Her work led to several invited talks at GreenTouch, Bell Labs, Optical Network Design and Modelling conference, Optical Fiber Communications conference, International Conference on Computer Communications, EU Future Internet Assembly, IEEE Sustainable ICT Summit and IEEE 5G World Forum and collaboration with Nokia and Huawei.

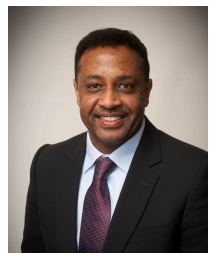

**Jaafar M. H. ELmirghani** is Fellow of IEEE, Fellow of the IET, Fellow of the Institute of Physics and is the Director of the Institute of Communication and Power Networks and Professor of Communication Networks and Systems within the School of Electronic and Electrical Engineering, University of Leeds, UK. He joined Leeds in 2007 having been full professor and chair in Optical Communications at the University of Wales Swansea 2000-2007. He received the BSc in Electrical Engineering, First Class Honours from the University of Khartoum in 1989 and was awarded all 4 prizes in the department for academic distinction. He received the PhD in the synchronization of optical systems and optical receiver design from the University of Huddersfield UK in 1994 and the DSc in Communication Systems and Networks from University of Leeds, UK, in 2012. He co-authored Photonic Switching Technology: Systems and Networks, (Wiley) and has published over 550 papers. He was Chairman of the IEEE UK and RI Communications Chapter and was Chairman of IEEE Comsoc Transmission Access and Optical Systems Committee and Chairman of IEEE Comsoc Signal Processing and Communication Electronics (SPCE) Committee. He was a member of IEEE ComSoc Technical Activities Council' (TAC), is an editor of IEEE Communications Magazine and is and has been on the technical program committee of 41 IEEE ICC/GLOBECOM conferences between 1995 and 2020 including 19 times as Symposium Chair. He was founding

Chair of the Advanced Signal Processing for Communication Symposium which started at IEEE GLOBECOM'99 and has continued since at every ICC and GLOBECOM. Prof. Elmirghani was also founding Chair of the first IEEE ICC/GLOBECOM optical symposium at GLOBECOM'00, the Future Photonic Network Technologies, Architectures and Protocols Symposium. He chaired this Symposium, which continues to date. He was the founding chair of the first Green Track at ICC/GLOBECOM at GLOBECOM 2011, and is Chair of the IEEE Sustainable ICT Initiative, a pan IEEE Societies Initiative responsible for Green ICT activities across IEEE, 2012-present. He has given over 90 invited and keynote talks over the past 15 years. He received the IEEE Communications Society 2005 Hal Sobol award for exemplary service to meetings and conferences, the IEEE Communications Society 2005 Chapter Achievement award, the University of Wales Swansea inaugural 'Outstanding Research Achievement Award', 2006, the IEEE Communications Society Signal Processing and Communication Electronics outstanding service award, 2009, a best paper award at IEEE ICC'2013, the IEEE Comsoc Transmission Access and Optical Systems outstanding Service award 2015 in recognition of "Leadership and Contributions to the Area of Green Communications", the GreenTouch 1000x award in 2015 for "pioneering research contributions to the field of energy efficiency in telecommunications", the IET 2016 Premium Award for best paper in IET Optoelectronics, shared the 2016 Edison Award in the collective disruption category with a team of 6 from GreenTouch for their joint work on the GreenMeter, the IEEE Communications Society Transmission, Access and Optical Systems technical committee 2020 Outstanding Technical Achievement Award for outstanding contributions to the "energy efficiency of optical communication systems and networks". He was named among the top 2% of scientists in the world by citations in 2019 in Elsevier Scopus, Stanford University database which includes the top 2% of scientists in 22 scientific disciplines and 176 sub-domains. He was elected Fellow of IEEE for "Contributions to Energy-Efficient Communications," 2021. He is currently an Area Editor of IEEE Journal on Selected Areas in Communications series on Machine Learning for Communications, an editor of IEEE Journal of Lightwave Technology, IET Optoelectronics and Journal of Optical Communications, and was editor of IEEE Communications Surveys and Tutorials and IEEE Journal on Selected Areas in Communications series on Green Communications and Networking. He was Co-Chair of the GreenTouch Wired, Core and Access Networks Working Group, an adviser to the Commonwealth Scholarship Commission, member of the Royal Society International Joint Projects Panel and member of the Engineering and Physical Sciences Research Council (EPSRC) College. He has been awarded in excess of £30 million in grants to date from EPSRC, the EU and industry and has held prestigious fellowships funded by the Royal Society and by BT. He was an IEEE Comsoc Distinguished Lecturer 2013-2016. He was PI of the £6m EPSRC Intelligent Energy Aware Networks (INTERNET) Programme Grant, 2010-2016 and is currently PI of the EPSRC £6.6m Terabit Bidirectional Multi-user Optical Wireless System (TOWS) for 6G LiFi, 2019-2024. He leads a number of research projects and has research interests in communication networks, wireless and optical communication systems.